\documentclass[
aps,
prb,
graphicx,
reprint,
]{revtex4-2}

\usepackage[utf8]{inputenc}
\usepackage[T1]{fontenc}
\usepackage{mathptmx}
\usepackage{etoolbox}
\usepackage{graphicx}
\usepackage{amsmath}
\usepackage{amssymb}
\usepackage{hyperref}
\usepackage{multirow}
\usepackage{xcolor}

\begin{document}

\title{Competition between phase ordering and phase segregation in the Ti$_x$NbMoTaW and Ti$_x$VNbMoTaW refractory high-entropy alloys}

\author{Christopher D. Woodgate}
\email[]{Christopher.Woodgate@warwick.ac.uk}
\affiliation{Department of Physics, University of Warwick, Coventry, CV4 7AL, United Kingdom}
\author{Julie B. Staunton}
\email[]{J.B.Staunton@warwick.ac.uk}
\affiliation{Department of Physics, University of Warwick, Coventry, CV4 7AL, United Kingdom}

\date{March 14, 2024}

\begin{abstract}
Refractory high-entropy alloys are under consideration for applications where materials are subjected to high temperatures and levels of radiation, such as in the fusion power sector. However, at present, their scope is limited because they are highly brittle at room temperature. One suggested route to mitigate this issue is by alloying with Ti. In this theoretical study, using a computationally efficient linear-response theory based on density functional theory calculations of the electronic structure of the disordered alloys, we study the nature of atomic short-range order in these multi-component materials, as well as assessing their overall phase stability. Our analysis enables direct inference of phase transitions in addition to the extraction of an atomistic, pairwise model of the internal energy of an alloy suitable for study via, {\it e.g.} Monte Carlo simulations. Once Ti is added into either the NbMoTaW or VNbMoTaW system, we find that there is competition between chemical phase ordering and segregation. These results shed light on observed chemical inhomogeneity in experimental samples, as well as providing fundamental insight into the physics of these complex systems.
\end{abstract}

\maketitle

\section{Introduction}

A promising class of high-entropy materials are the refractory high-entropy alloys, first reported by Senkov {\it et al.} in 2010~\cite{senkov_refractory_2010}. These alloys are systems where four or more refractory elements are alloyed in roughly equal ratios to form a single phase solid solution~\cite{senkov_development_2018}. The prototypical refractory high-entropy alloys are NbMoTaW and VNbMoTaW~\cite{senkov_refractory_2010, senkov_development_2018, senkov_mechanical_2011}, but numerous other compositions have been experimentally synthesised, such as MoNbHfZrTi~\cite{guo_microstructure_2015}, TaNbHfZrTi~\cite{senkov_microstructure_2011}, and HfMoNbTaTiZr~\cite{juan_enhanced_2015}. These materials often have enhanced physical properties compared to their base elements, and are of interest for advanced nuclear applications~\cite{pickering_high-entropy_2021} owing both to their exceptional mechanical properties at high-temperature~\cite{senkov_microstructure_2012} and their outstanding radiation resistance~\cite{el-atwani_outstanding_2019}. However, at room temperature, alloys such as NbMoTaW and VNbMoTaW are typically brittle~\cite{han_effect_2017, han_microstructures_2018} limiting their applicability. One suggested route for improvement is to alloy with Ti~\cite{han_effect_2017} but experiments find that chemical inhomogeneities emerge when these systems are processed~\cite{han_microstructures_2018, zhang_senary_2015}. In this work we explore this phenomenon via {\it ab initio} computational modelling.

Our approach is based on a perturbative analysis of the internal energy of the disordered solid solution evaluated via density functional theory (DFT) calculations. Here we study the effect of the addition of Ti on the two prototypical refractory high-entropy alloys, NbMoTaW and VNbMoTaW. In an earlier work~\cite{woodgate_short-range_2023} where the addition of Ti was not considered, we found that our modelling approach correctly predicts that the single phase solid solution for both NbMoTaW and VNbMoTaW is stable to comparatively low temperatures, with eventual B2 (for NbMoTaW) and B32 (for VNbMoTaW) chemical orderings emerging at 559~K and 750~K respectively. However, in the present study, we find that the addition of Ti leads to competition between chemical phase ordering and phase segregation in both of these systems. Furthermore, the temperatures at which chemical ordering/segregation are predicted to emerge are increased. These results go some way towards explaining the chemical inhomogeneities evident in experimental samples. Moreover we propose that our accurate, computationally-efficient modelling approach can accelerate the exploration of the composition space of high-entropy refractory alloys to find new compositions with desirable physical properties and to reduce the number of `trial and error' material syntheses and characterisations.

The paper is organised as follows. Section~\ref{sec:methodology} outlines the details of our modelling approach. The computational analysis not only enables us to infer order-disorder transitions directly, but also facilitates extraction of simple, pairwise atomistic models suitable for further study via, {\it e.g.} Monte Carlo simulations. Section~\ref{sec:results} presents results for the Ti$_x$NbMoTaW and Ti$_x$VNbMoTaW systems in the region $0\leq x \leq 1$ and examines the complex predicted phase behaviour of these systems. Because our modelling approach is based on DFT calculations, we also elucidate the origins of chemical ordering in terms of the underlying electronic structure of these complex materials. Finally, in Section~\ref{sec:conclusions}, we summarise our key findings, give an outlook on their implications and outline potential further work.

\section{Methodology}
\label{sec:methodology}

Computational modelling approaches have a crucial role to play in the process of materials design, discovery, and optimisation. They can both facilitate understanding of the phase behaviour and physical properties of existing materials, as well as guide experiment by suggesting novel compositions and/or processing techniques for new ones. This is particularly important in the space of high-entropy alloys and materials, as the vast space of potential compositions makes large-scale searches experimentally challenging. A number of methods have been developed and successfully used to study the phase behaviour of high-entropy alloys. Examples include semi-empirical approaches such as CALPHAD~\cite{li_calphad-aided_2023} as well as those based on DFT calculations, including large-scale supercell studies~\cite{tamm_atomic-scale_2015} and fitted inter-atomic interactions~\cite{huhn_prediction_2013, kim_interaction_2023}, machine learned interatomic potentials~\cite{kostiuchenko_impact_2019}, and cluster expansions~\cite{fernandez-caballero_short-range_2017}. Another class of DFT-based methodologies are based on effective medium theories such as the coherent potential approximation (CPA)~\cite{singh_atomic_2015, kormann_long-ranged_2017}.

\begin{figure}
    \centering
    \includegraphics{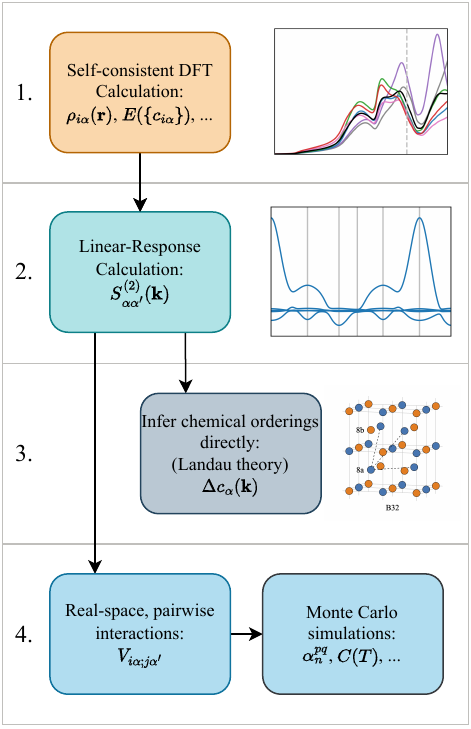}
    \caption{Visualisation of the workflow used in this paper for modelling the phase behaviour of a given multicomponent alloy, as discussed in Sec.~\ref{sec:methodology}.}
    \label{fig:workflow}
\end{figure}

Our approach for modelling the phase behaviour of multicomponent alloys falls into the last of these categories, and has been discussed extensively in earlier works~\cite{woodgate_compositional_2022, woodgate_short-range_2023, woodgate_interplay_2023, woodgate_atomic_2023, woodgate_integrated_nodate}, so we only outline the key details of the theory here. The methodology is complementary to other DFT-based modelling techniques, and is based on earlier work on binary alloys~\cite{gyorffy_concentration_1983, staunton_compositional_1994}. The workflow, a schematic of which is shown in Fig.~\ref{fig:workflow} can be broken down into four key steps:
\begin{enumerate}
    \item {\bf Electronic Structure.} A self-consistent DFT calculation is performed to model the electronic structure (and associated internal energy) of the disordered solid solution. This is performed within the Korringa-Kohn-Rostoker (KKR) formulation of DFT~\cite{ebert_calculating_2011, faulkner_multiple_2018}, using the CPA to model the electronic structure of the disordered solid solution~\cite{soven_coherent-potential_1967, johnson_total-energy_1990}.
    \item {\bf Perturbative Analysis.} Using the $S^{(2)}$ theory for multicomponent alloys~\cite{khan_statistical_2016}, which represents a linear response theory for the KKR-CPA internal energy of the disordered solid solution, we perform a perturbative analysis to understand the dominant atom-atom correlations in the disordered phase.
    \item {\bf Landau Theory.} Application of a Landau-type linear response theory to an approximate form of the free energy of the disordered solid solution, where the variation of the internal energy comes from the perturbative analysis, we are able to predict both the temperature of the initial chemical disorder-order transition as well as its nature in terms of ordered structures and (partial) atomic site occupancies.
    \item {\bf Monte Carlo Simulation.} From the linear response calculation, we can extract a simple, pairwise atomistic model of the internal energy of the system, which enables lattice-based Monte Carlo simulations to be performed to explore the phase space of the system in more detail.
\end{enumerate}
The above workflow therefore facilitates a thorough analysis of the phase behaviour of a given system. Although our modelling approach is lattice-based, ongoing work demonstrates that the on-lattice configurations obtained via Monte Carlo simulations can be relaxed and/or deformed for use in subsequent supercell studies, for example by training machine-learned interatomic potentials~\cite{shenoy_collinear-spin_nodate}. We now discuss each of the above methodological steps in more detail.

\subsection{Electronic Structure: The Internal Energy of the Solid Solution}

Given an underlying Bravais lattice, the configuration of an alloy can be specified by a set of \emph{site occupancies}, $\{\xi_{i\alpha}\}$, where $\xi_{i\alpha}=1$ if site $i$ is occupied by an atom of species $\alpha$, and $\xi_{i\alpha}=0$ otherwise. The constraint that every lattice site is occupied by one (and only one) atom is expressed as 
\begin{equation}
    \sum_\alpha \xi_{i\alpha} = 1.
    \label{eq:occupancy_condition}
\end{equation} 
We note that vacancies can also be treated in this formalism by considering them as a separate chemical species.

For a given configuration (and a sufficiently small system), the internal energy associated with a configuration, $E_\text{int}[\{\xi_{i\alpha}\}]$, can be evaluated directly via DFT calculations~\cite{martin_electronic_2004}. However, such calculations are computationally demanding and this renders direct evaluation of the partition function and associated thermodynamic quantities challenging. 

In this work, therefore, we will use an alternative description of the configuration of the system by working directly with the ensemble average of the site-wise configurations, the so called \emph{site-wise concentrations}. These are partial atomic site occupancies defined as
\begin{equation}
    c_{i\alpha} := \langle \xi_{i\alpha} \rangle,
    \label{eq:site-wise_concentrations}
\end{equation}
where $\langle \cdot \rangle$ denotes an ensemble average. Note that by Eq.~\ref{eq:occupancy_condition}, we have that $0\leq c_{i\alpha} \leq 1$. These site-wise concentrations represent order parameters classifying potential chemically ordered phases.

In the limit of high-temperature, where the alloy is disordered, these quantities become spatially homogeneous, {\it i.e.} an atom can occupy any lattice site with equal probability. This is equivalent to the statement that
\begin{equation}
    \lim_{T \to \infty} c_{i\alpha} = c_\alpha,
\end{equation}
where $c_\alpha$ is the overall (total) concentration of species $\alpha$. It remains to evaluate the average internal energy of a system with these homogeneous site occupancies, written $\langle E_\text{int} \rangle [\{c_{i\alpha}\}]$. Such a scheme is provided by the CPA~\cite{soven_coherent-potential_1967, johnson_total-energy_1990} within the KKR formulation of DFT~\cite{faulkner_multiple_2018}. The CPA constructs an effective medium of electronic scatterers whose average scattering properties approximate those of the disordered alloy, and it has been shown to reproduce successfully many physical properties of disordered systems. For example, KKR-CPA calculations have previously been shown to successfully reproduce the smeared out Fermi surface of the CrCoFeNi high-entropy alloy~\cite{robarts_extreme_2020}, as well as a variety of magnetic~\cite{billington_bulk_2020} and transport~\cite{jin_tailoring_2016} properties.

\subsection{Perturbative Analysis and Landau Theory: The $S^{(2)}$ Theory for Multicomponent Alloys}

To assess the energetic cost of perturbations to the disordered solution, we begin with an expression for its Landau free energy, $\Omega$. In general, this takes the form
\begin{equation}
    \Omega = U - TS -\mu N,
\end{equation}
where $U$ is the internal energy, $T$ the temperature, $S$ the entropy, $\mu$ the chemical potential(s), and $N$ the number(s) of particles in the system. For the description of the alloy considered in this paper, the free energy is approximated via
\begin{equation}
    \Omega^{(1)} = \langle E_\text{int} \rangle [\{c_{i\alpha}\}] - \beta^{-1} \sum_{i \alpha} c_{i\alpha} \ln{c_{i\alpha}} - \sum_{i \alpha} \nu_{i \alpha} c_{i\alpha}.
\end{equation}
In the above expression, the first term represents the average internal energy as obtained within the CPA, the second term represents the so-called entropy of mixing, and the third term represents chemical potentials which, in principle, can vary for each chemical species and lattice site. The chemical potentials serve as Lagrange multipliers in the theory and conserve overall concentrations of each chemical species.

We then make a Landau series expansion of this free energy about the spatially homogeneous reference state. Writing the inhomogeneous site occupancies as a perturbation to the homogeneous system,
\begin{equation}
    c_{i\alpha} = c_\alpha + \Delta c_{i \alpha},
    \label{eq:perturbation}
\end{equation}
this series expansion takes the form
\begin{equation}
\begin{aligned}
    \Omega^{(1)}[\{c_{i\alpha}\}] =& \Omega^{(1)}\left[ \{ c_\alpha \} \right] + 
\sum_{i\alpha} 
\frac{\partial \Omega^{(1)}}{\partial c_{i\alpha}}\bigg\rvert_{\{c_\alpha \}} \Delta c_{j\alpha} 
\\ &+
\frac{1}{2}\sum_{\substack{i\alpha \\ j\alpha'}}
\frac{\partial^2 \Omega^{(1)}}{\partial c_{i\alpha} \partial
c_{j\alpha'}}\bigg\rvert_{\{c_\alpha \}}
 \Delta c_{i\alpha} \Delta c_{j\alpha'} + \dots
\end{aligned}
\label{eq:landau}
\end{equation}
Owing to the homogeneity of the high-temperature reference state, and because we impose the condition that chemical fluctuations must conserve the overall concentrations of each species, the first-order term vanishes~\cite{khan_statistical_2016}.

To second order, the change in free energy due to a chemical perturbation is therefore written
\begin{equation}
    \Delta \Omega^{(1)} = 
\frac{1}{2}\sum_{\substack{i\alpha \\ j\alpha'}}
\frac{\partial^2 \Omega^{(1)}}{\partial c_{i\alpha} \partial
c_{j\alpha'}}\bigg\rvert_{\{c_\alpha \}}
 \Delta c_{i\alpha} \Delta c_{j\alpha'}.
\label{eq:second-order_real}
\end{equation}
The variation of the chemical potentials is not considered relevant to the underlying physics~\cite{khan_statistical_2016}, so their derivatives are set to zero and the second derivative of the free energy can therefore be expressed as
\begin{equation}
\frac{\partial^2 \Omega^{(1)}}{\partial c_{i\alpha} \partial
c_{j\alpha'}}  = \frac{\partial^2 \langle E_\text{int}\rangle}
{\partial c_{i\alpha} \partial c_{j\alpha'}} -\beta^{-1}
\left(\delta_{ij}\delta_{\alpha\alpha'}\frac{1}{c_{j\alpha'}}\right).
\label{eq:second-derivative_real}
\end{equation}
It can be shown that these second derivatives relate directly to an estimate of the two-point correlation function, {\it i.e.} the atomic short-range order~\cite{woodgate_compositional_2022, khan_statistical_2016}. For notational convenience, we define a new quantity $S^{(2)}_{i\alpha;j\alpha'}$ via
\begin{equation}
    S^{(2)}_{i \alpha; j \alpha'} := \frac{\partial^2 \langle E_\text{int} \rangle_0}
{\partial c_{i\alpha} \partial c_{j\alpha'}}.
\end{equation}
A scheme by which to evaluate these derivatives by considering infinitesimal perturbations to the CPA reference medium has been considered in detail in Ref.~\citenum{khan_statistical_2016} and first implemented in Ref.~\cite{woodgate_compositional_2022}. The same implementation is used here.

\begin{figure}[b!]
    \includegraphics[width=0.49\linewidth]{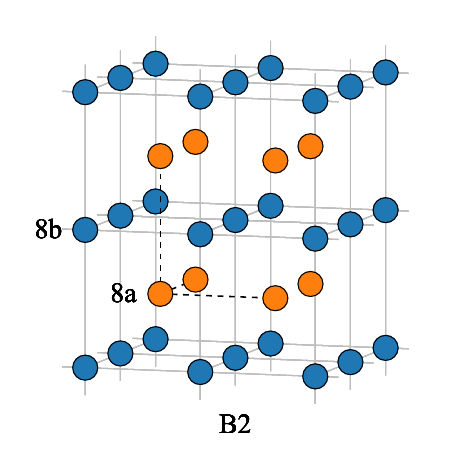}
    \includegraphics[width=0.49\linewidth]{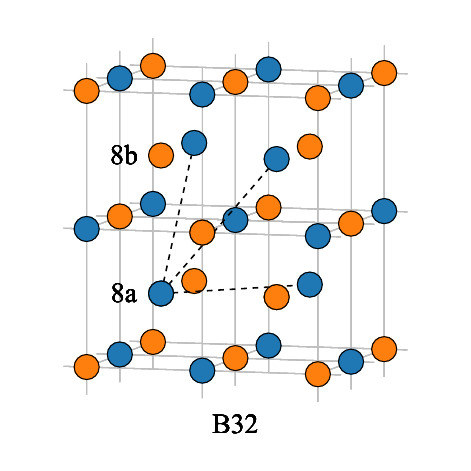}
\caption{Visualisations of the B2 and B32 ordered structures imposed on the bcc lattice. A B2 ordering in an equiatomic binary system is described by a wavevector $\mathbf{k}_\text{us} = (0,0,1)$ and equivalent, with chemical polarisation $\Delta c_\alpha = 1/\sqrt{2} (1, -1)$. The B32 ordering is described by a concentration wave with the same chemical polarisation, but this time with a wavevector of $\mathbf{k}_\text{us} = (1/2, 1/2, 1/2)$ and equivalent.}
\label{fig:structures}
\end{figure}

Due to the underlying crystal lattice, it is convenient to work with Fourier-transformed variables in a so-called concentration wave formalism as pioneered by Khachaturyan~\cite{khachaturyan_ordering_1978} and Gyorffy and Stocks~\cite{gyorffy_concentration_1983}. The site-wise concentrations are written
\begin{equation}
c_{i\alpha} = c_\alpha + \sum_{\mathbf{k}} e^{i \mathbf{k} \cdot
\mathbf{R}_i} \Delta c_{\alpha}(\mathbf{k}),
\end{equation}
where $\mathbf{k}$ is a wavevector, $\{\mathbf{R}_i\}$ are the positions of the lattice sites, and $\Delta c_{i\alpha}(\mathbf{k})$ are the concentration waves. Eqs.~\ref{eq:second-order_real} and \ref{eq:second-derivative_real} are then combined and written in reciprocal space as 
\begin{equation}
    \Delta \Omega^{(1)} = \frac{1}{2} \sum_\mathbf{k} \sum_{\alpha\alpha'}
 \Delta c_{\alpha}(\mathbf{k}) \left[
\beta^{-1} \frac{\delta_{\alpha \alpha'}}{c_\alpha} - S^{(2)}_{\alpha\alpha'}(\mathbf{k})
\right] \Delta c_{\alpha'}(\mathbf{k}).
\label{eq:chemical_stability}
\end{equation}
The term in square brackets is referred to as the \emph{chemical stability matrix}. Above a disorder-order transition temperature, the eigenvalues of this matrix are positive for all $\mathbf{k}$. However with decreasing temperature, we expect that at some temperature $T_\text{us}$, where `us' is shorthand for unstable, the lowest-lying eigenvalue of this matrix will pass through zero for some wavevector $\mathbf{k}_\text{us}$. The associated chemical ordering is described by the eigenvector $\Delta c_\alpha (\mathbf{k})$. For example, for a binary alloy, the B2 ordering visualised in Fig.~\ref{fig:structures} is described by a wavevector $\mathbf{k}_\text{us} = (0,0,1)$ and equivalent, with a \emph{chemical polarisation} $\Delta c_\alpha = 1/\sqrt{2} (1, -1)$. (Convention is that the chemical polarisation be normalised to be a unit vector.) Similarly, a B32 ordering is described by a concentration wave with the same chemical polarisation, but this time with a wavevector of $\mathbf{k}_\text{us} = (1/2, 1/2, 1/2)$ and equivalent. Finally, the case of phase segregation is described by a wavevector of $\mathbf{k}_\text{us} = (0, 0, 0)$, representing a concentration wave of infinite length.

\subsection{Monte Carlo Simulations}

In addition to the Landau theory used to infer chemical orderings directly, it is possible to map the results of the $S^{(2)}$ calculation back to a simple, pairwise atomistic model, the Bragg-Williams model~\cite{bragg_effect_1934, bragg_effect_1935}, which works directly with the discrete site-occupancies. The Hamiltonian for the Bragg-Williams model takes the form
\begin{equation}
H\left( \{ \xi_{i\alpha} \} \right) 
    = \frac{1}{2} \sum_{i,j} \sum_{\alpha \alpha'} V_{i\alpha;j\alpha'} \, \xi_{i\alpha} \xi_{j\alpha'}.
\label{eq:bragg_williams_hi}
\end{equation}
Given a Hamiltonian of this form, it can be shown that $V_{i\alpha;j\alpha'} = -S^{(2)}_{i \alpha; j \alpha'}$, making the $S^{(2)}$s an unambiguous best choice of parameter to use in this model. This model is suitable for study via lattice-based Metropolis Monte Carlo simulations using Kawasaki dynamics~\cite{landau_guide_2014} ({\it i.e.} only permitting swaps of atoms) to conserve overall concentrations of each species. The algorithm picks two lattice sites at random and computes the change in energy, $\Delta H$, realised by swapping the site occupancies. If $\Delta H < 0$ the move is accepted unconditionally, while if $\Delta H \geq 0$ the move is accepted with probability $e^{-\beta \Delta H}$. This is repeated until equilibrium is achieved. 

Once a simulation has been equilibrated at a given temperature, we extract two key quantities of interest. To quantify atomic short-range order in our simulations, we use the Warren-Cowley atomic short-range order parameters~\cite{cowley_approximate_1950, cowley_short-range_1965}, denoted $\alpha^{pq}_n$ and defined as
\begin{equation}
    \alpha^{pq}_n = 1 - \frac{P^{pq}_n}{c_q},
\end{equation}
where $n$ refers to the $n$th coordination shell, $P^{pq}_n$ is the conditional probability of an atom of type $q$ neighbouring an atom of type $p$ on coordination shell $n$, and $c_q$ is the overall concentration of atom type $q$. When $\alpha^{pq}_n>0$, $p$-$q$ pairs are disfavoured on shell $n$, while when $\alpha^{pq}_n<0$ they are favoured. The value 0 corresponds to the ideal, random, solid solution. 

The second quantity of interest is a measure of the configurational contribution to the specific heat capacity of the system, estimated via the fluctuation-dissipation theorem. At thermodynamic equilibrium, this theorem allows us to estimate the specific heat capacity (SHC) as
\begin{equation}
    C = \frac{1}{k_b T^2} \left( \langle E^2 \rangle - \langle E \rangle^2 \right),
\end{equation}
to obtain our SHC curves. A combined plot of the Warren-Cowley order parameters and specific heat capacity as a function of temperature for a given simulation facilitates understanding of the phase behaviour of a given system. In addition, it is possible to use sample configurations drawn from these lattice-based Monte Carlo simulations as inputs to other modelling approaches to predict materials properties.

\begin{figure*}[t!]
\includegraphics[width=\textwidth]{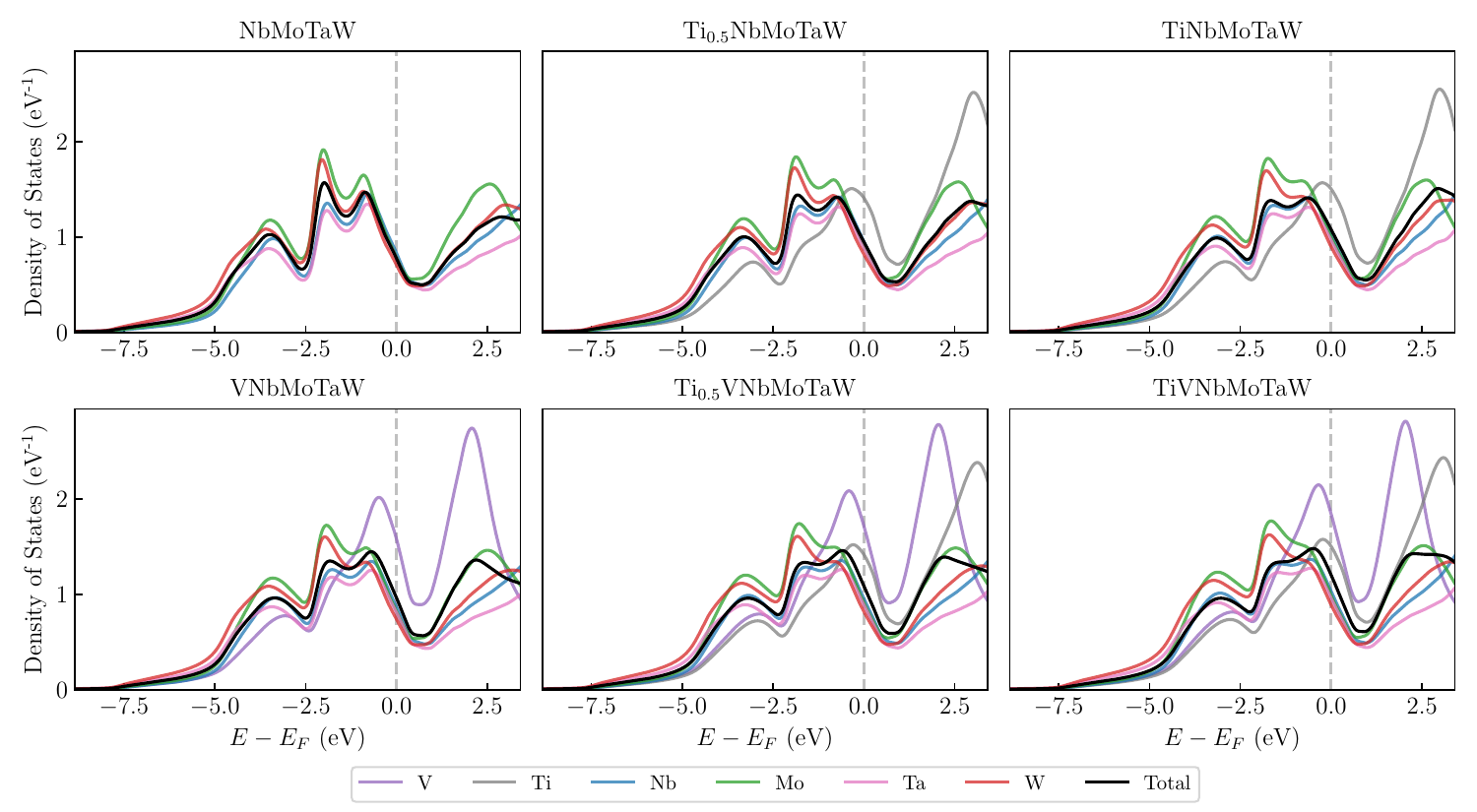}
\caption{Comparison of the total and species-resolved density of states for different values of $x$ for the disordered solid solution modelled within the coherent potential approximation for both Ti$_x$NbMoTaW and Ti$_x$VNbMoTaW. The Fermi level, $E_F$, is indicated by a vertical, grey, dashed line. The $3d$ elements V and Ti have species-resolved curves which differ substantially in character from the $4d$/$5d$ elements Nb, Mo, Ta, and W, suggesting the potential for stronger ordering tendencies.}
\label{fig:dos_comparison}
\end{figure*}

\begin{figure*}[t!]
\includegraphics[width=\textwidth]{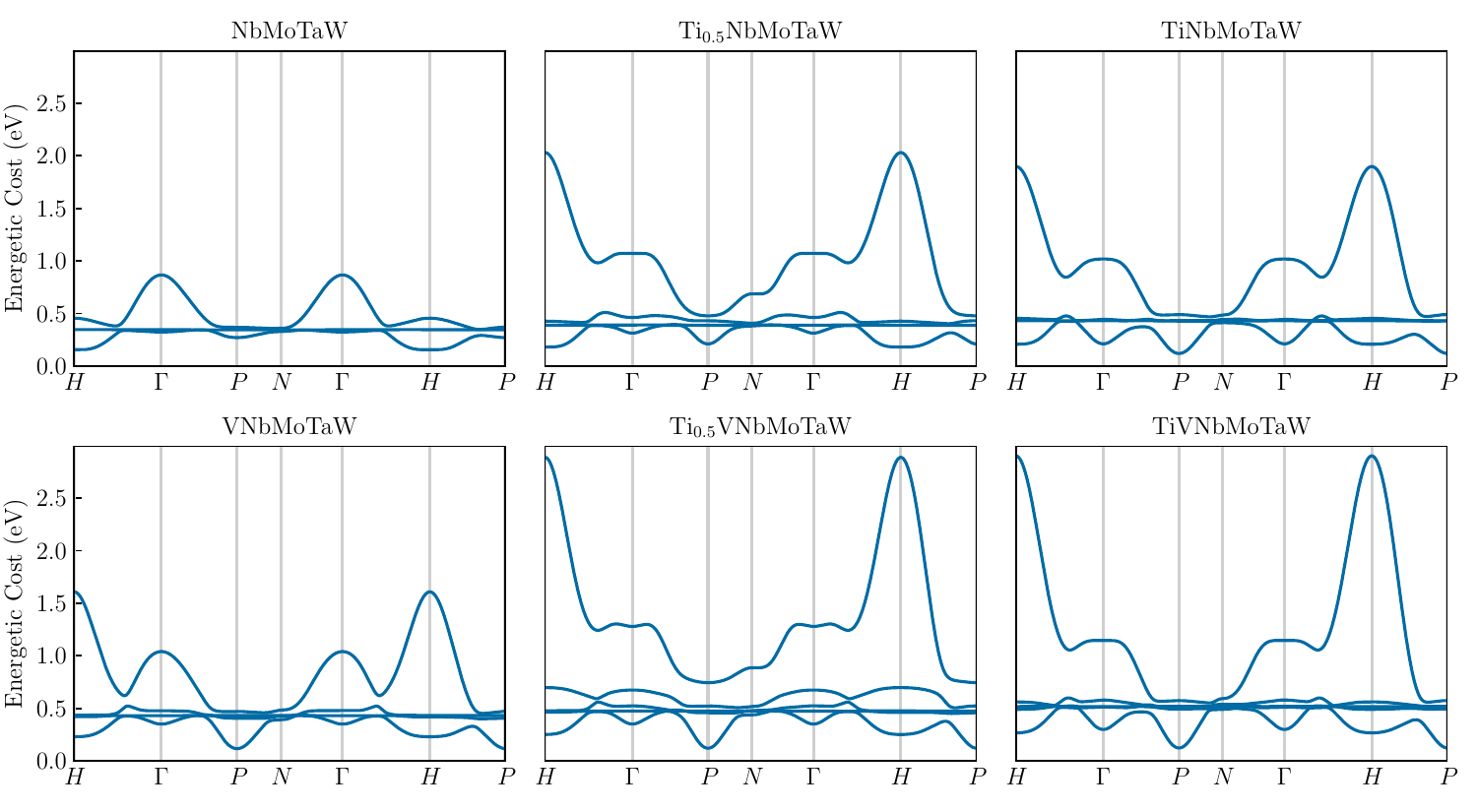}
\caption{Plots of the eigenvalues of the chemical stability matrix, $[\beta^{-1} \Psi^{-1}(\mathbf{k})]_{\alpha \alpha'}$, around the irreducible Brillouin Zone of the body-centred cubic lattice, evaluated at $T=1000$~K. Without the addition of Ti, the minima lie at $H$ and $P$, indicative of B2 and B32 ordering, respectively. However, with increasing Ti concentration, for both systems, an additional minimum at $\Gamma$, indicative of phase segregation, can be seen to emerge.}
\label{fig:linear_response}
\end{figure*}

\section{Results and Discussion}
\label{sec:results}

\subsection{Electronic Structure}

We begin by performing a self-consistent DFT calculation to model the electronic structure of the disordered solid solution. We use the all-electron HUTSEPOT code~\cite{hoffmann_magnetic_2020} to construct the self-consistent potentials of the KKR-CPA formulation of DFT. We perform spin-polarised, scalar-relativistic calculations within the atomic sphere approximation (ASA)~\cite{stocks_complete_1978}, employing an angular momentum cutoff of $l_\text{max} = 3$ for basis set expansions, a $20\times20\times20$ $\mathbf{k}$-point mesh for integrals over the Brillouin zone, and a 24 point semi-circular Gauss-Legendre grid in the complex plane to integrate over valence energies. We use the local density approximation and the exchange-correlation functional is that of Perdew-Wang~\cite{perdew_accurate_1992}. bcc lattice parameters for NbMoTaW and VNbMoTaW are set at 3.226~\AA~and 3.183~\AA~respectively, consistent with their experimental values~\cite{han_microstructures_2018, senkov_refractory_2010}. For TiNbMoTaW and TiVNbMoTaW ({\it i.e.} the case $x=1$) these values are set at 3.240~\AA~and 3.209~\AA~respectively, again consistent with experimentally determined values~\cite{han_effect_2017}. For intermediate values of $x$ we interpolate linearly between these using Vegard's law~\cite{denton_vegards_1991}.

Previous CALPHAD modelling has suggested the possibility of the emergence of an hcp phase at low temperatures with increasing Ti concentration~\cite{han_microstructures_2018}. To investigate this aspect, we perform a transformation to the hcp structure conserving the overall volume-per-atom and compute the difference in total energy per atom between the bcc and hcp structures as a function of $x$, the results of which are visualised in the Supplemental Material~\cite{supplemental}. Within the ASA, using the CPA to average over disorder, we find that the bcc structure is consistently favoured over the hcp structure in the region $0\leq x \leq 2$, {\it i.e.} up to and beyond the range of $x$ considered in this paper. We therefore only consider the alloy on a bcc lattice for the remainder of this paper.

Proceeding, in Fig.~\ref{fig:dos_comparison} we visualise the total and species-resolved density of states (DoS) for both systems for three indicative values of $x$. The total DoS is given by the weighted average of the species-resolved curves. It can be seen that pairs of chemical species which are isoelectronic, {\it e.g.} Nb/Ta and Mo/W, have species-resolved curves which lie almost on top of one another, which we expect to lead to weak correlations between these pairs of elements. Compared to the $4d$ and $5d$ elements, the two $3d$ elements considered here, V and Ti, can be seen to have narrower $d$ bands with significantly different profiles around $E_F$, which we expect to lead to strong ordering tendencies between these and other elements, as noted in our earlier study~\cite{woodgate_short-range_2023}.

Another important feature to note is the charge-transfer between elements. In an alloy system where there is an atomic size discrepancy, where the lattice parameter sits somewhere between that of the pure elements (approximately in accordance with Vegard's law), there is often a transfer of charge from the large atoms to the smaller ones, corresponding to the charge density associated with the `large' atom spilling over into the Winger-Seitz cell associated with the `small' atom. In our calculations, we find that this effect is generally strongest for V and Ta, with V gaining charge and Ta losing it. For example, in the five-component VNbMoTaW, V (proton number 23) has an associated average charge per atom of 23.177~$e$, while Ta (proton number 73) has an associated charge per-atom of 72.901~$e$. This is consistent with V being the `smallest' atom and Ta one of the `largest' for the systems considered here.

\subsection{Perturbative Analysis}
\label{sec:perturbative_results}

\begin{table*}[t!]
\begin{ruledtabular}
\begin{tabular}{lccrrrrrr}
System     & $\mathbf{k}$-point & Eigenvalue (eV) & $\Delta c_\text{Ti}$ & $\Delta c_\text{V}$ & $\Delta c_\text{Nb}$ & $\Delta c_\text{Mo}$ & $\Delta c_\text{Ta}$ & $\Delta c_\text{W}$ \\ \hline
TiNbMoTaW  & $H$                & 0.290           & $-0.107$             &                     & $-0.361$             & $0.636$              & $-0.553$             & $0.386$             \\
           & $\Gamma$           & 0.297           & $-0.784$             &                     & $0.168$              & $-0.163$             & $0.508$              & $0.271$             \\
           & $P$                & 0.227           & $-0.828$             &                     & $-0.011$             & $0.101$              & $0.246$              & $0.493$             \\ \hline
TiVNbMoTaW & $H$                & 0.368           & $-0.252$             & $0.167$             & $-0.361$             & $0.589$              & $-0.530$             & $0.388$             \\
           & $\Gamma$           & 0.402           & $-0.611$             & $-0.331$            & $0.138$              & $-0.144$             & $0.589$              & $0.360$             \\
           & $P$                & 0.253           & $-0.605$             & $-0.487$            & $0.099$              & $0.172$              & $0.314$              & $0.508$            
\end{tabular}
\end{ruledtabular}
\caption{Comparison of eigenvalues and their associated chemical polarisations for Ti$_x$NbMoTaW and Ti$_x$VNbMoTaW for $x=1$. All are evaluated at $T=1000$ K. For both systems, for the special points $H$, $\Gamma$, and $P$ (indicative of B2 ordered, phase separation, and B32 ordering, respectively), the eigenvalues are close in value, indicative of competition between different concentration wave modes. At the $\Gamma$ point, the chemical polarisation is dominated by Ti, suggesting a tendency for Ti to segregate away from other elements.}\label{table:sample_eigenvalues}
\end{table*}

\begin{table*}[t!]
\begin{ruledtabular}
\begin{tabular}{lcccrrrrrr}
System         & Ti Concentration ($x$) & $T_\text{us}$ (K) & $\mathbf{k}_\text{us}$ ($\frac{2 \pi}{a}$) & $\Delta c_\text{Ti}$ & $\Delta c_\text{V}$ & $\Delta c_\text{Nb}$ & $\Delta c_\text{Mo}$ & $\Delta c_\text{Ta}$ & $\Delta c_\text{W}$ \\ \hline
Ti$_x$NbMoTaW  & 0                      & 557               & $(0,0,1)$                                  &                      &                     & $-0.421$             & $0.566$              & $-0.568$             & $0.424$             \\
               & 0.5                    & 584               & $(1/2, 1/2, 1/2)$                           & $-0.725$             &                     & $-0.168$             & $0.047$              & $0.217$              & $0.629$             \\
               & 1                      & 673               & $(1/2, 1/2, 1/2)$                          & $-0.828$             &                     & $-0.011$             & $0.101$              & $0.246$              & $0.493$             \\ \hline
Ti$_x$VNbMoTaW & 0                      & 698               & $(1/2, 1/2, 1/2)$                          &                      & $-0.828$            & $-0.005$             & $0.090$              & $0.247$              & $0.495$             \\
               & 0.5                    & 726               & $(1/2, 1/2, 1/2)$                          & $-0.397$             & $-0.656$            & $0.054$              & $0.159$              & $0.298$              & $0.542$             \\
               & 1                      & 709               & $(1/2, 1/2, 1/2)$                          & $-0.605$             & $-0.487$            & $0.099$              & $0.172$              & $0.324$              & $0.507$            
\end{tabular}
\end{ruledtabular}
\caption{Predicted ordering temperatures ($T_\text{us}$), wavevectors ($\mathbf{k}_\text{us}$), and chemical polarisations for the Ti$_x$NbMoTaW and Ti$_x$VNbMoTaW systems for $x=0$, 0.5, 1. The addition of Ti generally increases predicted ordering temperatures on account of strong correlations between Ti and other elements.}\label{table:ordering_temperatures}
\end{table*}

Proceeding, we perform a linear response calculation to assess the dominant atomic short- and long-range order in these systems, as outlined in Section~\ref{sec:methodology}. Visualised in Fig.~\ref{fig:linear_response} are eigenvalues of the chemical stability matrix around the irreducible Brillouin zone (IBZ) of the bcc lattice, evaluated at a temperature of {$T=1000$~K}. Note that, for an $s$-species alloy, there are $s-1$ eigenvalues due to the constraint that the overall concentration of each species be conserved.

To interpret these \emph{concentration wave modes}, there are three key features to consider. The first is the shape of the modes, as these tell us about the nature and strength of atom-atom interactions in the disordered phase. Strongly varying modes are associated with strong interactions, while weakly varying modes are associated with chemical species interacting weakly. The second feature is the location of the minima, as these tell us about the dominant correlations in the disordered phase, and to infer the likely chemical ordering. Finally, it is necessary to know the \emph{chemical polarisation} of a mode, {\it i.e.} $\Delta c_\alpha$ to understand the chemical species to which it relates. Some sample eigenvalues, and chemical polarisations at a few high-symmetry points of the IBZ for TiNbMoTaW and TiVNbMoTaW are provided in Table~\ref{table:sample_eigenvalues}, while the remaining data can be found in the repository associated with this publication.

Considering first the Ti$_x$NbMoTaW system, the top row of Fig.~\ref{fig:linear_response}, we see that for the case $x=0$ we have three clear modes emerging. The first is a mode dipping at $H$ and peaking at $\Gamma$, which is associated with B2 ordering tendencies and, in concentration space, is polarised to suggest that Nb and Ta will sit on one sublattice, while Mo and W will sit on the other. The two other modes present are near-flat, and are associated with the isoelectronic pairs Nb-Ta and Mo-W, which are very weakly correlated in our calculations. On adding Ti by increasing $x$, however, we see the emergence of a new, strongly varying mode, introducing competing minima at $P$ and $\Gamma$, associated with B32 ordering and phase segregation respectively. Notably, the chemical polarisation of the concentration wave mode at the $\Gamma$ point, as shown in Table~\ref{table:sample_eigenvalues}, suggests Ti and Mo segregating from Nb, Ta and W, which is entirely consistent with the experimentally observed segregation in Ref.~\cite{han_microstructures_2018}. These results therefore confirm that the addition of Ti produces competing interactions in this system, with the potential for a variety of chemical orderings and/or phase segregation.

Proceeding, we move on to the results for the Ti$_x$VNbMoTaW alloy, {\it i.e.} the bottom row of Fig.~\ref{fig:linear_response}. For the case $x=0$, there is an additional mode compared to NbMoTaW, which dips at $P$ and peaks at $H$, and is associated with a B32 ordering between V and the other elements present. This was analysed in detail in our earlier work~\cite{woodgate_short-range_2023} and is associated with the large charge-transfer between V and the other elements present. Again, upon the addition of Ti, however, a new, strongly varying mode materialises and introduces a competing minimum at the $\Gamma$ point, indicative of phase segregation. As shown in Table~\ref{table:sample_eigenvalues}, this minimum has a chemical polarisation suggesting Ti, V, and Mo segregating away from Nb, Ta, and W, again consistent with the experimental data~\cite{zhang_senary_2015, han_effect_2017}. As before, we interpret these data as suggesting that Ti does not mix well with the other elements present and introduces competing chemical interactions in the system.

For each of the considered systems, via application of the Landau theory outlined in Sec.~\ref{sec:methodology}, we estimate a transition temperature by computing the temperature for which the lowest-lying eigenvalue passes through zero. We also provide its chemical polarisation and associated wavevector, as these elucidate the nature of the ordered phase. These results are tabulated in Table~\ref{table:ordering_temperatures}. In summary for both systems, increasing Ti concentration leads to higher predicted ordering temperatures, and alters the chemical polarisation and/or wavevector describing the chemical ordering. It is evident that Ti drives strong atom-atom correlations in these systems. 

{ Both the temperature and nature of our predicted B2 ordering in NbMoTaW and predicted B32 ordering in VNbMoTaW are in good agreement with other DFT-based computational studies~\cite{huhn_prediction_2013, fernandez-caballero_short-range_2017, kostiuchenko_impact_2019, kim_interaction_2023}, including those using GGA functionals, and we take this as evidence that the LDA adequately captures the relevant physics in these systems. We also note that our results are not hugely sensitive to the choice of lattice parameter. For example, for the five-component VNbMoTaW, for a 2\% decrease in lattice parameter, we predict a slightly increased ordering temperature of 740 K, while for a 2\% increase in lattice parameter, we predict a slightly decreased ordering temperature of 627 K. (This effect is the origin of the small numerical difference between ordering temperatures predicted in this study for VNbMoTaW and NbMoTaW, using experimental lattice parameters, and the ordering temperatures obtained in Ref.~\cite{woodgate_short-range_2023}, using LDA-optimised lattice parameters.) Despite the modest changes in ordering temperature, the predicted chemical ordering is still into the B32 structure, and the chemical polarisations remain very numerically close to those predicted at the experimental lattice parameter. These findings therefore verify the robustness of our results.}

\begin{figure*}[t!]
\includegraphics[width=\textwidth]{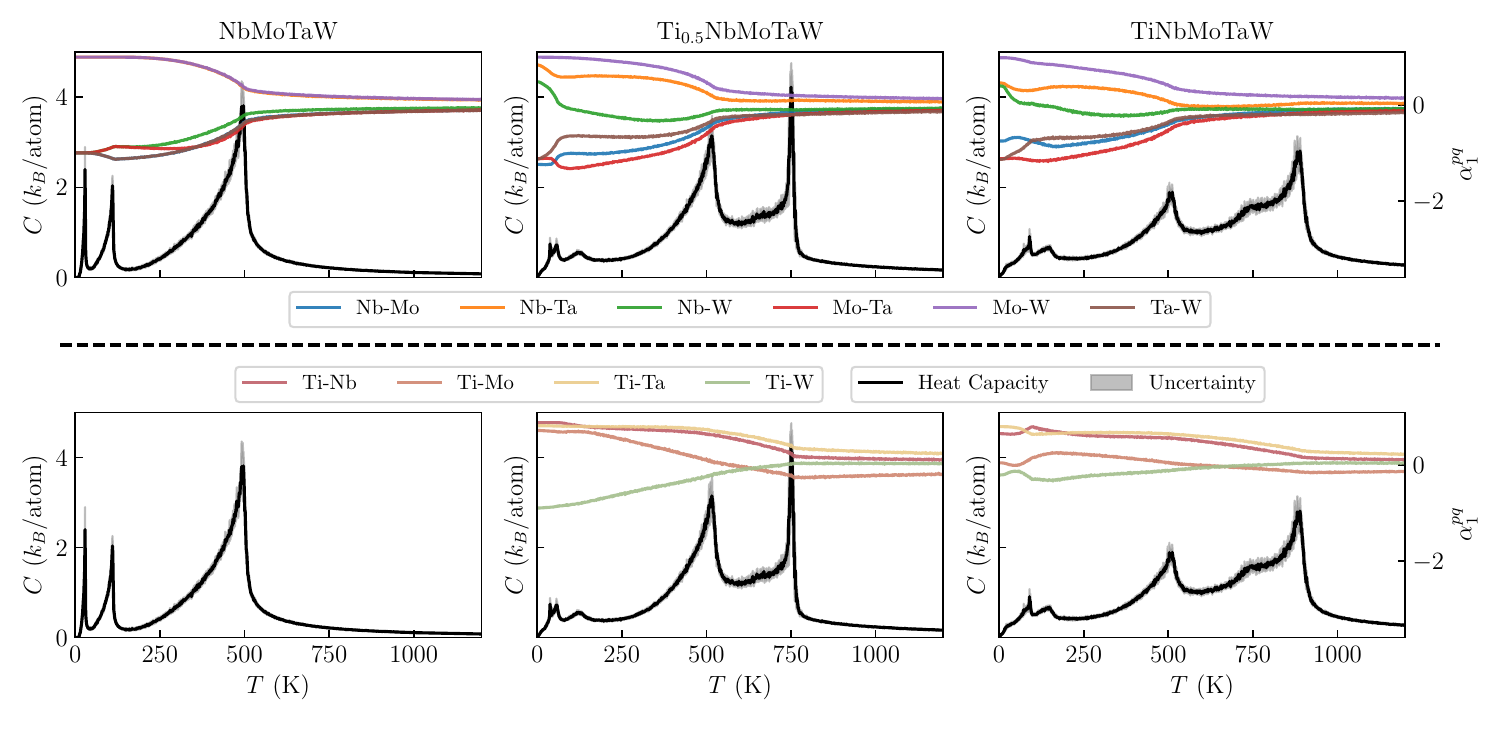}
\caption{Plots of the Warren-Cowley ASRO parameters, $\alpha_{n}^{pq}$, and specific heat capacity, $C$, for Ti$_x$NbMoTaW as a function of temperature, calculated from an ensemble of lattice-based Monte Carlo simulations. The top row shows atom-atom correlations excluding Ti, while the bottom row shows correlations between Ti and other elements. The addition of Ti can be seen to introduce an additional Ti-driven ordering compared to the NbMoTaW system.}
\label{fig:warren-cowley_tinbmotaw}
\end{figure*}

\begin{figure*}[t!]
\includegraphics[width=\textwidth]{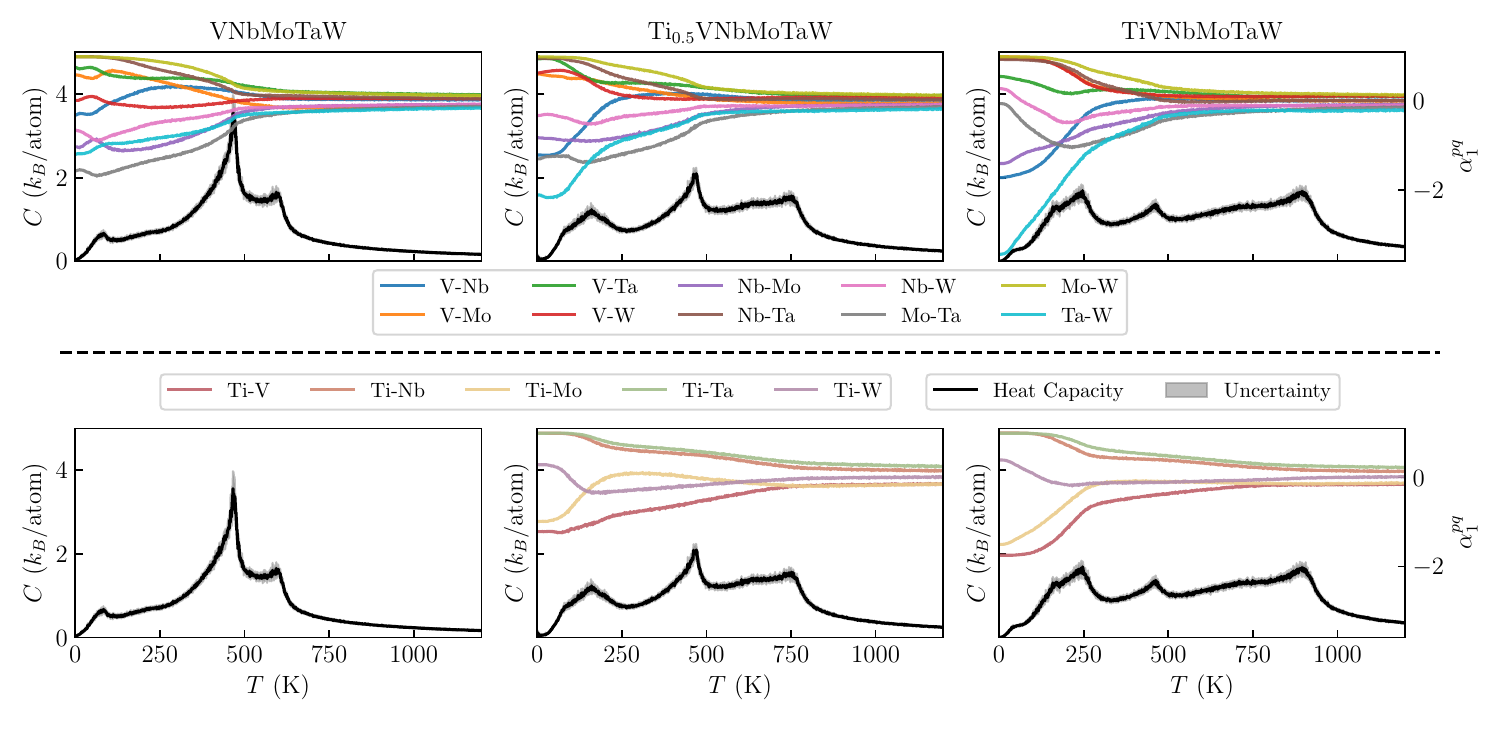}
\caption{Plots of the Warren-Cowley ASRO parameters, $\alpha_{n}^{pq}$, and specific heat capacity, $C$, for Ti$_x$VNbMoTaW  as a function of temperature, calculated from an ensemble of lattice-based Monte Carlo simulations. The top row shows atom-atom correlations excluding Ti, while the bottom row shows correlations between Ti and other elements. As for the Ti$_x$NbMoTaW system of Fig.~\ref{fig:warren-cowley_tinbmotaw}, the addition of Ti can be seen to introduce an additional Ti-driven ordering compared to the VNbMoTaW system.}
\label{fig:warren-cowley_tinvbmotaw}
\end{figure*}

\subsection{Monte Carlo Simulations}

From the linear response calculations performed in reciprocal space, we are able to backwards Fourier transform and recover a pairwise, real-space interaction, as described in Section~\ref{sec:methodology}. Our fitted interactions are provided in the database associated with this study { and are tabulated in the Supplemental Material~\cite{supplemental}}. We find that a fit to the first four coordination shells of the bcc lattice captures the reciprocal space data with acceptable accuracy{, and that interactions are strongest on the first two coordination shells}. Typically, we find that the strongest interactions are between Ti, V, and the other elements present, while $4d/5d$ pairs such as Nb-Ta and Mo-W, which are isoelectronic, interact weakly. { Our fitted interactions exhibit clear dependence on system composition, which reflects the importance of a full treatment of the multicomponent system rather than extrapolating interactions from binary subsystems. However, typically, the sign and order of magnitude of atom-atom pair interactions remains consistent across the range of Ti concentrations considered.} It should be noted that the linear response results of Sec.~\ref{sec:perturbative_results} include the Onsager correction of Ref.~\cite{khan_statistical_2016}, which serves to restore important on-site sum rules concerning atomic short-range order and charge and drives down estimated transition temperatures. However, the simple pairwise form of Eq.~\ref{eq:bragg_williams_hi} does not permit inclusion of an Onsager correction directly. Inclusion of the Onsager correction in the atomistic simulations is a subject of ongoing study.

Proceeding, we performed lattice-based Monte Carlo simulations (simulated annealing) of both the Ti$_x$NbMoTaW and Ti$_x$VNbMoTaW systems for $x=0$, 0.5, 1. All simulation cells consisted of a system of $16\times 16 \times 16$ cubic bcc unit cells for a total of 8192 atoms in the simulation cell. At each temperature, following an initial burn-in period to achieve equilibrium, statistics were gathered over a run of $10^4$ Monte Carlo steps per atom. All data were averaged across an ensemble of 10 simulations, which enables an extraction of an uncertainty on key quantities such as the specific heat capacity of the simulation.

Visualised in Figures~\ref{fig:warren-cowley_tinbmotaw} and ~\ref{fig:warren-cowley_tinvbmotaw} are plots of the Warren-Cowley ASRO parameters at nearest-neighbour distance and specific heat capacity as a function of temperature for Ti$_x$NbMoTaW and Ti$_x$VNbMoTaW respectively. (Warren-Cowley ASRO parameters at second-nearest neighbour distance are provided in the supplementary material~\cite{supplemental}.) The top row of panels for each figure shows Warren-Cowley parameters excluding Ti, while the bottom row shows the Warren-Cowley parameters for correlations between Ti and the other elements present.

Considering first the results for Ti$_x$NbMoTaW (Fig.~\ref{fig:warren-cowley_tinbmotaw}), we see that there is a clear trend of increasing transition temperature with increasing Ti concentration, as the initial peak in SHC is moved to higher temperatures. In the NbMoTaW system, as discussed in our earlier work~\cite{woodgate_short-range_2023}, the initial peak in the SHC is associated with a B2 ordering, with Nb and Ta atoms preferentially sitting on one sublattice, and Mo and W atoms preferentially sitting on the other. { This predicted B2 ordering is in good agreement with other computational studies of this system~\cite{huhn_prediction_2013, kormann_long-ranged_2017, kostiuchenko_impact_2019, kim_interaction_2023}. The peak in SHC associated with the B2 ordering} is maintained as Ti is added to the system, but a new, additional peak at higher temperatures emerges, associated with correlations between Ti and the other elements present. In particular, in our simulations, Ti favours pairing with W, and avoids Mo, Ta, and Nb. These results are consistent with the earlier linear response analysis suggesting that the introduction of Ti produces competing interactions and eventual phase segregation in this system.

Moving on to the results for the senary Ti$_x$VNbMoTaW system (Fig.~\ref{fig:warren-cowley_tinvbmotaw}), we see a similar picture emerging. Without Ti present, there is known to be emergent B32-like ordering in this system~\cite{fernandez-caballero_short-range_2017, woodgate_short-range_2023}, which is detected in both our linear response calculation and Monte Carlo simulations, indicated here by the peak in SHC between 500 and 750~K. As for the Ti$_x$NbMoTaW system, however, the introduction of Ti produces an additional peak in the SHC curves at higher temperature, which is dominated by correlations between Ti and the other elements present. In particular, our modelling suggests that, at nearest-neighbour distance, Ti-W pairs are favoured, while Ti-V pairs are disfavoured. %{ With the exception of the quarternary NbMoTaW system, with decreasing temperature all systems eventually exhibit complex, multiphase behaviour. Some indicative configurations are provided in the Supplemental Material~\cite{supplemental}.}

\section{Conclusions}
\label{sec:conclusions}

In summary, we have used a perturbative analysis based on DFT calculations of the internal energy of the disordered solid solution to examine atomic ordering tendencies in the Ti$_x$NbMoTaW and Ti$_x$VNbMoTaW refractory high-entropy alloys. From the perturbative analysis, we have fitted a pairwise, real-space interaction and explored the phase space further using Monte Carlo simulations. We have also discussed the origins of the dominant atom-atom correlations in terms of the materials' underlying electronic structure.

When Ti is not present ({\it i.e.} the case $x=0$) it is found that both NbMoTaW and VNbMoTaW form single-phase solid solutions down to relatively low temperatures, with predicted disorder-order transition temperatures of 557~K and 698~K respectively. The predicted transitions for these systems are chemical orderings, B2 and B32 respectively, and no significant phase segregation is expected. These results are consistent with both experimental and theoretical literature, as well as with our own earlier study~\cite{woodgate_short-range_2023}. 

However, with increasing Ti concentration ({\it i.e.} the case $x>0$) it is found that strong atom-atom correlations emerge in the system. This leads to competition between phase segregation and phase ordering, with the perturbative analysis suggesting that Ti and Mo (and, to a lesser extent, V) tends to segregate from Nb, Ta, and W. This effect is amplified with increasing Ti concentration.

These results shed light on the complex phase behaviour of these technologically relevant high-entropy alloys, as well as giving insight into the physical origins of the dominant atom-atom correlations in the solid solution, thus demonstrating the efficacy of this methodology. In future, we hope that our computationally efficient modelling approach can be used for further exploration of the space of high-entropy alloys and materials, and guide experiments towards new compositions with desirable physical properties. \\

\begin{acknowledgments}
C.D.W. was supported by a studentship within the UK Engineering and Physical Sciences Research Council-supported Centre for Doctoral Training in Modelling of Heterogeneous Systems, Grant No. EP/S022848/1. Computing facilities were provided by the
Scientific Computing Research Technology Platform of the University of Warwick.
\end{acknowledgments}

\section*{Author Declarations}

\subsection*{Conflict of Interest}

The authors have no conflicts to disclose.

\subsection*{Author Contributions}

{\bf Christopher D. Woodgate:} Conceptualization; Methodology (equal); Software (equal); Validation; Formal Analysis; Investigation; Data curation; Writing -- original draft; Writing -- review \& editing (equal); Visualisation. {\bf Julie B. Staunton:} Methodology (equal); Software (equal); Resources; Writing -- review \& editing (equal); Supervision; Project administration; Funding acquisition.

\section*{Data Availability Statement}

The data that support the findings of this study are openly available on Zenodo via the DOI \href{https://doi.org/10.5281/zenodo.10580621}{10.5281/zenodo.10580621}. The all-electron HUTSEPOT DFT code used for the KKR-CPA calculations is available at \href{https://hutsepot.jku.at/}{https://hutsepot.jku.at}. The $S^{(2)}$ code used for performing the linear-response calculation of the internal energy of the disordered solid solution is available from the authors upon reasonable request. The code for performing the lattice-based Monte Carlo simulations is available on Zenodo via the DOI \href{https://doi.org/10.5281/zenodo.10371917}{10.5281/zenodo.10371917}.

%\bibliography{ti_xvnbmotaw}


\end{document}